\documentclass[twoside]{article}
\usepackage[a4paper]{geometry}
\usepackage[utf8]{inputenc} 
\usepackage[OT1]{fontenc}
\usepackage{RR}
\usepackage{hyperref}

\newcommand\proof{\noindent\textit{Proof.}~}
\newcommand\myendproof{\xspace$\Box$\\\indent}
\def\isRR{}

\RRNo{8522}
\RRdate{April 2014}
\RRauthor{Venmugil~Elango\thanks[osu]{Ohio State Univ.}
\and Fabrice Rastello\thanks{Inria, Univ. de Grenoble-Alpes}
\and Louis-No\"el~Pouchet\thanks{Univ. of California--Los Angeles}
\and J.~Ramanujam\thanks{Louisiana State University}
\and P.~Sadayappan\thanksref{osu}
}
\authorhead{Elango, Rastello, Pouchet, Ramanujam, and Sadayappan}
\RRetitle{On Characterizing the Data Movement Complexity of Computational DAGs for Parallel Execution}

\RRtitle{Caractérisation de la complexité I/O d'une application sur architecture distribuée.} 

\RRabstract{

Technology trends are making the cost of data movement increasingly
dominant, both in terms of energy and time, over the cost of
performing arithmetic operations in computer systems.  The fundamental
ratio of aggregate data movement bandwidth to the total computational
power (also referred to the \emph{machine balance parameter}) in
parallel computer systems is decreasing. It is therefore of
considerable importance to characterize the inherent data movement
requirements of parallel algorithms, so that the minimal architectural
balance parameters required to support it on future systems can be
well understood.

In this paper, we develop an extension of the well-known red-blue
pebble game to develop lower bounds on the data movement complexity
for the parallel execution of computational directed acyclic graphs
(CDAGs) on parallel systems. We model multi-node multi-core parallel
systems, with the total physical memory distributed across the nodes
(that are connected through some interconnection network) and in a
multi-level shared cache hierarchy for processors within a node.  We
also develop new techniques for lower bound characterization of
non-homogeneous CDAGs. We demonstrate the use of the methodology by
analyzing the CDAGs of several numerical algorithms, to develop lower
bounds on data movement for their parallel execution.

\comment{
Architectural advances over the last few decades has significantly
increased the gap between computation and communication costs, causing
the data movement to account for the majority of energy expenditure
and execution time. 
The advent of multicore and distributed-memory systems, with
hierarchical memories, has led to data transfers between different
processors, and also between different memory levels within a
processor.
Fundamental characterization of these data movements for various
algorithms will be an important metric for comparing algorithms.
In this paper, we revisit the problem of data access complexity, first
addressed by the seminal work of \hk~\cite{hong.81.stoc} using the
formalism of the red/blue pebble game. Our paper makes the following
contributions: (1) we extend the red/blue pebble game model to the
multicore/distributed-memory machines with hierarchical memories, (2)
we provide the lower bounds on the data movement complexity for
different algorithms, and (3) we provide the corresponding upper-bounds
to measure the tightness of the lower bounds.
}
}
\RRresume{Avec les technologies actuelles, le coût d'une communication (autant en terme de temps que d'énergie) devient de plus en plus dominant face au coût de calcul. Le ratio entre bande passante et puissance de calcul (\emph{machine balance parameter} en anglais) dans les systèmes parallèles ne fait que décroître. Il est donc fondamental de savoir caractériser la complexité d'une application en terme de nombre de mouvements de données minimal.

Cet article développe une extension du jeu des pions rouges et noirs (\emph{red-blue pebble game}) afin de dériver des bornes inférieures sur la complexité de communication d'un graphe de tâches acyclique (CDAG) dans un environnement de calcul distribué. Les systèmes modélisés à cet effet sont des multi-coeurs à mémoire distribuées, ainsi que des multi-coeurs à mémoire partagée. Les techniques développées s'appliquent à des CDAG non homogènes. 

Nous illustrons la méthodologie à travers l'analyse de CDAGs de plusieurs algorithmes, dérivant ainsi des bornes inférieures sur leur complexité de communication sur machines distribuées.
} 
\RRkeyword{red-blue pebble game, I/O complexity, I/O lower bound, Computational DAG, parallel architectures, data movement} 
\RRmotcle{jeux des pions rouges et noirs, complexité I/O, nombre minimal de communications, graphe de tâches acyclique, architectures distribuées, communications} 

\RRprojets{GCG}
\URRhoneAlpes

\usepackage{setspace}
\usepackage{multirow}
\usepackage{amsmath,amssymb}
\usepackage{paralist}
\usepackage{alltt,url}
\usepackage{xspace}
\usepackage{epsfig}
\usepackage{pstricks,graphics}
\usepackage{IEEEeqnarray,progSTF}
\usepackage{listings}
\usepackage{float}
\usepackage{epsfig}
\usepackage{pstricks,graphics}
\usepackage{enumerate}
\usepackage{caption}
\usepackage{amsmath,amssymb}
\usepackage{balance}
\usepackage{amsmath,amssymb}
\usepackage{alltt,url}
\usepackage{xspace}
\usepackage{epsfig}
\usepackage{pstricks,graphics}

\usepackage{balance}

\usepackage{amsmath,amsfonts,amsbsy} 
\usepackage{fancyvrb,keyval,ifthen}
\usepackage{graphicx}
\graphicspath{{./}{logos/}}

\def\DO    {{\bf do}}
\def\ENDDO {{\bf end do}}
\def\UNTIL {{\bf until}}
\def\FOR   {{\bf for}}

\newtheorem{lemma}{Lemma}
\newtheorem{theorem}{Theorem}
\newtheorem{definition}{Definition}

\newtheorem{corollary}{Corollary}

\newcommand{\mat}[1]{#1}
\newcommand{\vctr}[1]{#1}
\newcommand{\norm}[1]{\left\lVert #1 \right\rVert_2}
\newcommand{\inprod}[1]{\left\langle\langle #1 \right\rangle\rangle}
\DeclareMathOperator*{\argmin}{arg\,min}

\newcommand\hk{Hong \& Kung\xspace}

\def\C{\textit{C}\xspace} 
\def\P{{\cal P}\xspace}
\def\Q{\textit{Q}\xspace}

\def\S{\textit{S}\xspace}

\def\IO{\textit{IO}\xspace} 
\newcommand\Hmin[1]{\textit{H}(#1)\xspace}
\newcommand\Umax[1]{\textit{U}(#1)\xspace}

\def\Sset{\mathcal{S}}
\def\Tset{\mathcal{T}}
\def\V{V\xspace}

\newcommand\card[1]{\left|#1\right|}

\newcommand\Min[1]{\textsf{Min}(#1)}
\newcommand\In[1]{\textsf{In}(#1)}
\newcommand\Out[1]{\textsf{Out}(#1)}

\newcommand\Succ[1]{\textsf{Desc}(#1)}
\newcommand\Pred[1]{\textsf{Anc}(#1)}
\newcommand\Doms[1]{\textsf{Dom}(#1)}

\newcommand*{\rom}[1]{\uppercase\expandafter{\romannumeral #1\relax}}

\long\def\comment#1{}

\begin{document}

\makeRR

\section{Introduction}
Recent technology trends have resulted in much greater rates of
improvement in computational processing rates of processors than the
bandwidths for data movement across nodes or within the memory/cache
hierarchies within nodes in a parallel system.  This mismatch between
maximum computational rate and peak memory bandwidth means that data
movement and communication costs of an algorithm will be increasingly
dominant determinants of performance.  Although hardware techniques
for data pre-fetching and overlapping of computation with
communication can alleviate the impact of memory access latency on
performance, the mismatch between maximum computational rate and peak
memory bandwidth is much more fundamental; \emph{the only solution is
  to limit the total rate of data movement between components of a
  parallel system to rates that can be sustained by the interconnects
  at different components and levels of a parallel computer system.}

It is therefore of considerable importance to develop techniques to
characterize lower bounds on the data movement complexity of parallel
algorithms. We address this problem in this paper. We formalize the
problem by developing a parallel extension of the red-blue pebble game
model introduced by Hong and Kung in their seminal work
~\cite{hong.81.stoc} on characterizing the data access complexity
(called \emph{I/O complexity} by them) for sequential execution of
computational directed acyclic graphs (CDAGs). Our extended pebble
game abstracts data movement in scalable parallel computers today,
which are comprised of multiple nodes interconnected by a
high-bandwidth interconnection network, with each node containing a
number of cores that share a hierarchy of caches and the node's
physical main memory. 

In contrast to some other prior efforts that have modeled lower bounds
for data movement in parallel computations, we focus on relating data
movement lower bounds to the critical architectural balance parameter
of the ratio of peak data movement bandwidth (in GBytes/sec) to peak
computational throughput (in GFLOPs) at different levels of a parallel
system.  We develop techniques for deriving lower bounds for data movement
for CDAGs under the parallel red-blue pebble game, and use the
techniques to analyze a number of numerical algorithms. Interesting
insights are provided on architectural bottlenecks that limit the
performance of the algorithms.

This paper makes several contributions:
\begin{compactitem}
\item It develops an extension of the red-blue pebble game that
  effectively models essential characteristics of scalable parallel
  computers with multi-level parallelism; (i) multiple nodes with local
  physical memory that are interconnected via a high-speed
  interconnection network like Infiniband or a custom interconnect
  (e.g., IBM BlueGene system \cite{BGRef}, or Cray XE6
  \cite{Cray-Ref}), and (ii) many cores at each node, that share a
  hierarchy of caches and the node's physical main memory.
\item It develops a lower bound analysis methodology that is effective
  for analysis of non-homogeneous CDAGs using a decomposition approach.
\item It develops new parallel lower-bounds analysis for a number of
  numerical algorithms.
\item It presents insights into implications on different
  architectural parameters in order to achieve scalable parallel
  execution of the analyzed algorithms.
\end{compactitem}

\section{Background}
\label{sec:background}

\subsection{Computational Model}
The model of computation we use is a computational directed acyclic
graph (CDAG), where computational operations are represented as graph
vertices and the flow of values between operations is captured by
graph edges.  
Two important characteristics
of this abstract form of representing a computation are that (1) there
is no specification of a particular order of execution of the
operations: although the program executes the operations in a
specific sequential order, the CDAG abstracts the schedule of
operations by only specifying partial ordering constraints as edges in
the graph; (2) there is no association of memory locations with the
source operands or result of any operation.
\comment{
\begin{figure}[h!tb]
\begin{center}
\begin{minipage}{4cm}
\begin{verbatim}
for (i = 1; i < 4; ++i)
  S += A[i-1] + A[i];
\end{verbatim}
\hrule
\end{minipage}

\includegraphics[width=4cm]{./figures/example4_modif.pdf}
\end{center}
\vspace{-.3cm}
\caption{\label{fig:ex1} Example of a CDAG. Input vertices are represented in black, output vertices in grey.}
\end{figure}
}
\noindent We use the notation of
Bilardi \& Peserico~\cite{bilardi2001characterization} to formally describe
CDAGs. We begin with the model of CDAG used by \hk:
\begin{definition}[CDAG-HK]
~\\
A computational directed acyclic graph (CDAG) is a 4-tuple
$C = (I,V,E,O)$ of finite sets such that:
(1) $I \subset V$ is the input set and all its vertices have no incoming edges;
(2) $E \subseteq V \times V$ is the set of edges;
(3) $G = (V, E)$ is a directed acyclic graph;
(4) $V-I$ is called the operation set and all its vertices have one
or more incoming edges;
(5) $O \subseteq V$ is called the output set.
\end{definition}


\subsection{The Red-Blue Pebble Game}
\label{sec:rbpg}

\hk used this computational model in their seminal work
\cite{hong.81.stoc}.  The inherent I/O complexity of a CDAG is the
minimal number of I/O operations needed while optimally playing the
\emph{Red-Blue pebble game}. This game uses two kinds of
pebbles: a fixed number of red pebbles that represent the small fast
local memory (could represent cache, registers, etc.), and an
arbitrarily large number of blue pebbles that represent the large slow
main memory. Starting with blue pebbles on all inputs nodes in the CDAG, the
game involves the generation of a sequence of steps to finally produce
blue pebbles on all outputs. A game is defined as follows.

\begin{definition}[Red-Blue pebble game~\cite{hong.81.stoc}]
\label{def:rbpg}
~\\
Given a CDAG $C=(I,V,E,O)$ such that any vertex with no incoming (resp. outgoing) edge is an element of $I$ (resp. $O$),
\S red pebbles and an arbitrary number
of blue pebbles, with a blue pebble on each \textit{input} vertex.
A complete game is any sequence of steps using the following
rules that results in
a final state with blue pebbles on all \textit{output} vertices:
\begin{compactitem}
\item[\textbf{R1 (Input)}] A red pebble may be placed on any vertex that has a blue pebble (load from slow to fast memory),
\item[\textbf{R2 (Output)}] A blue pebble may be placed on any vertex
that has a red pebble (store from fast to slow memory),
\item[\textbf{R3 (Compute)}] If all immediate predecessors of a vertex of $V-I$
have red pebbles, a red pebble may be placed on that vertex (execution or ``firing'' of operation),
\item[\textbf{R4 (Delete)}] A red pebble may be removed from any vertex (reuse storage).
\end{compactitem}
\end{definition}
\noindent The number of I/O operations for any complete game is the
total number of moves using rules R1 or R2, i.e., the total number
of data movements between the fast and slow memories. The inherent I/O
complexity of a CDAG is the smallest number of such I/O operations
that can be achieved, among all possible valid red-blue pebble games
on that CDAG. The \emph{optimal} red-blue pebble game is a game
achieving this minimal number of I/O operations.

\subsection{S-partitioning for Lower Bounds on I/O Complexity}

This red-blue pebble game provides an operational definition for the
I/O complexity problem.  However, it is not practically feasible to
generate all possible valid games for large CDAGs.  \hk developed a
novel approach for deriving I/O lower bounds for CDAGs by
relating the red-blue pebble game to a graph partitioning problem
defined as follows.
\begin{definition}[\S-partitioning of CDAG~\cite{hong.81.stoc}]
\label{def:spart}
~\\
 Given a CDAG $\C$. An \S-partitioning of $\C$ is a
collection of $h$ subsets of $\V$ such that:
\begin{compactenum}
\item[\textbf{P1}] $\forall i\neq j,\ \V_i\cap \V_j = \emptyset$, and $\bigcup_{i = 1}^h \V_i = \V$
\item[\textbf{P2}] there is no circuit between subsets
\item[\textbf{P3}] $\forall i,~~\exists D\in \Doms{{\V}_i}~~\textrm{such that}~~\card{D} \le \S$
\item[\textbf{P4}] $\forall i,~~\card{\Min{{\V}_i}} \le \S$
\end{compactenum}
\noindent where a dominator set of $\V_i$, $D\in\Doms{{\V}_i}$ is a set of vertices such that any path from $I$ to a vertex in $\V_i$ contains some vertex in $D$; the minimum set of $\V_i$, $\Min{{\V}_i}$ is the set of vertices in
${\V}_i$ that have all its successors outside of ${\V}_i$; and $\card{\textit{Set}}$ is the
cardinality of the set $\textit{Set}$.
We say that there is a circuit between two sets $V_i$ and $V_j$,
   if there is an edge from any vertex in $V_i$ to a vertex in $V_j$
   and vice-versa.
\end{definition}

\hk showed a construction for a 2\S-partition of a CDAG,
corresponding to any complete red-blue pebble game on that CDAG using
\S red pebbles, with a tight relationship between the number of vertex sets
$h$ in the 2\S-partition and the number of I/O moves $q$ in the
pebble-game shown next.
\begin{theorem}[Pebble game, I/O and 2\S-partition~\cite{hong.81.stoc}]\label{thm.hk} 
Any complete calculation of the red-blue pebble game on a CDAG
using at most \S red pebbles is associated with a 2\S-partition of the CDAG
such that $ \S\times h \ge q \ge \S\times(h-1), $
where $q$ is the number of I/O moves in the game and $h$ is the number of
subsets in the 2\S-partition.
\end{theorem}


The tight association from the above theorem between any pebble game
and a corresponding 2\S-partition provides the following key lemma that
served as the basis for \hk's approach to deriving lower
bounds on the I/O complexity of CDAGs.

\begin{lemma}[Lower bound on I/O~\cite{hong.81.stoc}]
\label{lemma:hk}
Let $\Hmin{2\S}$ be the minimal number of vertex sets for any valid
$2\S$-partition of a given CDAG (such that any vertex with no incoming -- resp. outgoing -- edge is an element of $I$ -- resp. $O$). Then the minimal number \Q
of I/O operations for any valid execution of the CDAG is bounded by: 
$\Q \ge \S \times(\Hmin{2\S}-1)$
\end{lemma}
This key lemma has been useful in proving I/O lower bounds for several
CDAGs~\cite{hong.81.stoc} by reasoning about the maximal number of
vertices that could belong to any vertex-set in a valid 2\S-partition.

The following corollary is generally useful while deriving I/O lower
bounds through 2\S-partitioning.

\begin{corollary}
\label{cor:hk}
Let $\Umax{2\S}$ be the largest vertex-set of any valid
$2\S$-partition of a given CDAG $\C=(I,V,E,O)$. Let $V'=V\setminus I$. Then the minimal number \Q of I/O
operations for any valid execution of the CDAG is bounded by: $\Q \ge
\S \times\left(\frac{\card{V'}}{\card{\Umax{2\S}}}-1\right)$

\comment{
\proof
Let $\Hmin{2\S}$ be the minimal number of vertex sets for any valid
$2\S$-partition of $\C$.
Since size of any vertex set of any valid $2\S$-partition of $\C$ is
less than or equal to $\card{\Umax{2\S}}$, $\Hmin{2\S}$ is bounded
from below by the value $\card{V'}/\card{\Umax{2\S}}$. Hence, $\Q \ge
\S \times(\Hmin{2\S}-1) \ge \S
\times\left(\frac{\card{V'}}{\card{\Umax{2\S}}}-1\right)$.
\myendproof
}
\end{corollary}

\section{Parallel Red-Blue-White Pebble Game}
\label{sec:parallel-game}

Application codes are typically constructed from a number of
sub-computations using the fundamental composition mechanisms of
sequencing, iteration and recursion.
For instance, the conjugate gradient method, described in
Sec.~\ref{sec:cg-overview}, consists of sequence of sparse
matrix-vector product, vector dot-product and SAXPY operations, 
for every iteration.
Applying the I/O lower bounding techniques directly on the CDAG of
such composite application codes can produce very weak lower bounds.
For instance, consider the following code segment.

\ifdefined\isRR\newpage\fi

\begin{minipage}{0.95\columnwidth}
\begin{lstlisting}[basicstyle=\small,frame=single,mathescape,numbers=left]
Inputs: $\vctr{p},~\vctr{q},~\vctr{r},~\vctr{s}$: Vectors of size $N$
Output: $sum$: Scalar
$\mat{A} = \vctr{p}\times \vctr{q}^T$
$\mat{B} = \vctr{r}\times \vctr{s}^T$
$\mat{C} = \mat{A}\mat{B}$
$sum = \sum_{i=1}^{N}\sum_{j=1}^{N}C_{ij}$
\end{lstlisting}
\end{minipage}

The computational complexity of this calculation can be simply obtained
by adding together the computational costs of the constituent
steps, i.e., $N^2 + N^2 + 2N^3 + N^2$ arithmetic operations. In contrast,
the data movement complexity for this computation cannot so simply be
obtained by adding together the data movement lower bounds for the
individual steps.
Let us consider data movement costs in a two-level memory hierarchy
with unbounded main memory and a limited number of words ($S$) in
fast storage -- this might represent the number of registers in the processor,
or scratchpad memory or cache memory.
It is known \cite{hong.81.stoc,toledo.jpdc,BDHS11} that an asymptotic lower bound on data movement between (arbitrarily large) slow memory
and fast memory for matrix multiplication of $N \times N$ matrices is
$N^3/2 \sqrt{2S}$. An outer-product of two vectors of size $N$
requires $2N$ input operations from slow memory and output of the $N^2$
results back to slow memory, i.e., total I/O of $2N + N^2$, independent
of the fast memory capacity $S$. Similarly, the last step has a data movement
complexity of $N^2+1$ I/O operations between slow and fast memory. But
a lower bound on the data movement complexity of the total calculation 
cannot be obtained by simply adding together contributions for the steps.
It is not even possible to assert that the maximum among them is a
valid lower bound on the data movement complexity of the total
calculation.
The reason is that data from a previous step could possibly be passed to
a later step in fast storage without having to be stored in main memory.
With $4N+4$ fast memory locations, it is feasible to perform the above
computation with a total of only $4N+1$ I/O operations, $4N$ to bring in
the four input vectors into fast memory, and repeatedly recompute elements
of A and B to contribute to an element of C, and when ready, accumulate
it into sum. The I/O complexity of the composite multi-step computation
is thus lower than that of the matrix multiply step contained in it.
This motivates us to split the CDAG
based on individual sub-computations, determine the lower bound for
each sub-CDAG separately, and finally compose the result to obtain
the I/O lower bound of the whole computation. 
%
However, using the original
red/blue pebble game model of \hk, as elaborated below, it is not
feasible to analyze the I/O complexity of sub-computations and simply
combine them by addition.

The \hk red/blue pebble game model places blue pebbles on all CDAG vertices
without predecessors, since such vertices are considered to hold
inputs to the computation, and therefore assumed to start off in slow memory.
Similarly, all vertices without successors are considered to be outputs
of the computation, and must have blue pebbles at the end of the game.
If the vertices of a CDAG corresponding to a composite application are
disjointly partitioned
into sub-DAGs, the analysis of each sub-DAG
under the \hk red/blue pebble game model will require the initial placement
of blue pebbles on all predecessor-free vertices in the sub-DAG, and final
placement of blue pebbles on all successor-free vertices in the sub-DAG.
The optimal pebble game for each sub-DAG will require at least one load
(R1) operation for each input and a store (R2) operation for each output.
But in playing the red/blue pebble game on the full composite CDAG, clearly
it may be possible to pass values in a red pebble between vertices
in different sub-DAGs, so that the I/O complexity is less than the sum
of the I/O costs for the optimal games for each sub-DAG. In fact, it is not even
possible to assert that the maximum among the I/O lower bounds
for sub-DAGs of a CDAG is a valid lower bound for the composite
CDAG.

In order to enable such decomposition, a modified game called the Red-Blue-White
pebble game~\cite{LB-TR} was defined, with the following changes to the
\hk pebble game model (the Red-Blue-White
pebble game is formally defined in Sec.~\ref{sec:rbwpg}):
\begin{compactenum}
\item {\bf Flexible input/output vertex labeling:} Unlike the \hk
model, where all vertices without predecessors must be input vertices, and
all vertices without successors must be output vertices,
the RBW model allows flexibility in indicating which vertices are labeled as
inputs and outputs.
In the modified
variant of the pebble game, predecessor-free vertices that are not
designated as input vertices do not have an initial blue pebble
placed on them. However, such vertices are allowed to fire using
rule R3 at any time, since they do not have any predecessor nodes
without red pebbles. Vertices without successors that are not labeled
as output vertices do not require placement of a blue pebble at the
end of the game. However, all compute vertices in the CDAG are required
to have fired for any complete game.
\item {\bf Prohibition of multiple evaluations of compute vertices:}
  The RBW game disallows recomputation of values on the CDAG, i.e., each
  non-input vertex is only allowed to evaluate once using rule
  R3. Several other efforts
  \cite{BDHS11,BDHS11a,bilardi2001characterization,michele13,ranjan11.fft,savage.cc.95,savage.book,savage.options,cook.pebbling,toledo.jpdc,ranjan12.rpyr,ranjan12.vertex}
  have also imposed such a restriction on the pebble game model. While
  such a model is indeed more restrictive than the original \hk model,
  the restriction in the model
  enables the development of techniques to form tighter lower
  bounds \cite{LB-TR}. 
\end{compactenum}

\subsection{The Red-Blue-White Pebble Game}
\label{sec:rbwpg}

\begin{definition}[Red-Blue-White (RBW) pebble game]
\label{def:rbwpg}
Given a CDAG $\C=(I,V,E,O)$,
\S red pebbles and an arbitrary number
of blue and white pebbles, with a blue pebble
on each \textit{input} vertex,
a complete game is any sequence of steps using the following
rules that results
a final state with white pebbles on all vertices and blue pebbles on all \textit{output} vertices:
\begin{compactitem}
\item[\textbf{R1 (Input)}] A red pebble may be placed on any vertex that has a blue pebble; a white pebble is also placed along with the red pebble, unless
the vertex already has a white pebble on it.
\item[\textbf{R2 (Output)}] A blue pebble may be placed on any vertex
that has a red pebble.
\item[\textbf{R3 (Compute)}] If a vertex $v$ does not have a white pebble
and all its immediate predecessors have red pebbles on them, a red pebble along with a white pebble may be placed on $v$.
\item[\textbf{R4 (Delete)}] A red pebble may be removed from any vertex (reuse storage).
\end{compactitem}
\end{definition}

In the modified rules for the RBW game, all vertices are required to have a
white pebble at the end of the game, thereby ensuring that the entire CDAG is
evaluated. Non-input vertices without predecessors do not have an initial blue
pebble on them, but they are allowed to fire using rule R3 at any time -- since
they have no predecessors, the condition in rule R3 is trivially satisfied. But
if all successors of such a node cannot be fired while maintaining a red pebble,
``spilling'' and reloading using R2 and R1 is forced because the vertex cannot
be fired again using R3.

Definition~\ref{def:spart} is adapted to this new game so that
Theorem~\ref{thm.hk} and thus Lemma~\ref{lemma:hk} can hold for the
RBW pebble game.

\begin{definition}[\S-partitioning of CDAG -- RBW pebble game]
\label{def:newspart}

Given a CDAG $\C$. An \S-partitioning of $\C$ is a
collection of $h$ subsets of $\V-I$ such that:
\begin{compactenum}
\item[\textbf{P1}] $\forall i\neq j,\ \V_i\cap\V_j = \emptyset$, and $\bigcup_{i = 1}^h V_i = \V-I$
\item[\textbf{P2}] there is no circuit between subsets
\item[\textbf{P3}] $\forall i,~~\card{\In{{V}_i}} \le \S$
\item[\textbf{P4}] $\forall i,~~\card{\Out{{V}_i}} \le \S$
\end{compactenum}
\noindent where the input set of $V_i$, $\In{{V}_i}$ is the set of
vertices of $\V\setminus V_i$ that have at least one successor in $V_i$; the
output set of $V_i$, $\Out{V_i}$ is the set of vertices of $V_i$ also
part of the output set $O$ or that have at least one successor outside of ${\V}_i$.
\end{definition}

The proof of Theorem~\ref{thm.hk} under the RBW pebble game is provided
in~\cite{LB-TR}.

For (sub-)graphs
without input/output sets, the application of S-partitioning will
however lead to a trivial partition with all vertices in a single set
(e.g., $h = 1$). A careful tagging of vertices as virtual input/output
nodes will be required for better I/O complexity estimates, as
described below.

\subsection{Decomposition}
\label{subsec:decomposition}


Definition~\ref{def:rbwpg} allows the partitioning of a CDAG $\C$ into
sub-CDAGs $C_1, C_2, \dots, C_p$, to compute lower bounds on the I/O
complexity of each sub-CDAG $\IO(C_1), \IO(C_2), \dots, \IO(C_p)$
independently and simply add them to bound the I/O complexity of
$\C$. This is stated in the following decomposition theorem, whose
proof may be found in \cite{LB-TR}.

\begin{theorem}[Decomposition]
\label{thm:decomposition}
~\\
Let $C=(I,V,E,O)$ be a CDAG. Let $V_1, V_2, \dots, V_p$ be an
arbitrary (not necessarily acyclic) disjoint partitioning of $V$ ($i\neq j \Rightarrow V_i\cap V_j=\emptyset$ and $\bigcup_{1\leq i\leq p} V_i=V$) and $C_1, C_2, \dots, C_p$ be the induced partitioning of $C$ ($I_i=I\cap V_i$, $E_i=E\cap V_i\times V_i$, $O_i=O\cap V_i$).
Then $\sum_{1\leq i\leq p} \IO(C_i)\leq \IO(C)$. In particular, if
$Q_i$ is a lower bound on the I/O cost of $C_i$, then $\sum_{1\leq i\leq p}
Q_i$ is a lower bound on the I/O cost of $C$.
\end{theorem}

\comment{
\proof
Consider an optimal valid game $\cal P$ for $C$, with cost $Q=\IO(C)$. We define the cost of $\cal P$ restricted to $V_i$, denoted as $Q_{|V_i}$, as the number of $R1$ or $R2$ transitions in $\cal P$ that involve a vertex of $V_i$. Clearly $Q=\sum_{1\leq i\leq p} Q_{|V_i}$. We can build from $\cal P$, a valid game ${\cal P}_{|V_i}$ for $C_i$, of cost $Q_{|V_i}$. This will prove that $\IO(C_i)\leq Q_{|V_i}$, and thus  $\sum_{1\leq i\leq p} \IO(C_i)\leq \sum_{1\leq i\leq p} Q_{|V_i} = Q = \IO(C)$. ${\cal P}_{|V_i}$ is built from $\cal P$ as follows\footnote{We do the reasoning here for the red-blue-white pebble game. Similar reasoning can also be done for the red-blue pebble game.}: (1) for any transition in $\cal P$ that involves a vertex $v$ in $V_i$, apply this transition in  ${\cal P}_{|V_i}$; (2) delete all other transitions in $\cal P$. Conditions for transitions $R1$, $R2$, and $R4$ are trivially satisfied. Whenever a transition $R3$ on a vertex $v$ is performed in $\cal P$, all the predecessors of $v$ must have a red pebble on them. Since all transitions of $\cal P$ on the vertices of $V_i$ are maintained in ${\cal P}_{|V_i}$, when $v$ is executed in ${\cal P}_{|V_i}$, all its predecessor vertices must have red pebbles, enabling transition $R3$.
\myendproof
}

We state the following corollary and theorem, which are useful in
practice for deriving tighter lower bounds. The complete proofs can be
found in~\cite{LB-TR}.

{\def\deltaI{\textit{dI}}
\def\deltaO{\textit{dO}}
\begin{corollary}[Input/Output Deletion]
\label{cor:delete}
Let $C$ and $C'$ be two CDAGs: $\C'=(I\cup \deltaI,V\cup \deltaI\cup \deltaO,E',O\cup\deltaO)$, $\C=(I, V, E'\cap V\times V, O)$. Then $\IO(C')$ can be bounded by a lower bound of $\IO(C)$ as follows:
\begin{equation}
\label{eq:delete}
\IO(\C) + |\deltaI| + |\deltaO|\leq \IO(\C')
\end{equation}
\end{corollary}
\comment{
\proof
The proof involves a direct application of the decomposition theorem, with $V_1=\deltaI$, $V_2=V$, and $V_3=\deltaO$. Indeed, the I/O complexity of the CDAG made of only $\deltaI$ is exactly $|\deltaI|$ (transition $R1$ is necessary for each of them to place a white pebble on it); the I/O complexity of the CDAG made only of $\deltaO$ is exactly $|\deltaO|$ (transition $R2$ is necessary for each of them to place a blue pebble on them).
\myendproof
}

There are cases where separating input/output vertices leads to very weak lower
bounds. This happens when input vertices have high fan out
such as for matrix-multiplication: if
we consider the CDAG for matrix-multiplication and remove all input and output
vertices, we get a set of independent chains that can each be computed with no
more than 2 red pebbles. To overcome this problem, the following theorem
allows us to compare the I/O of two CDAGs: a CDAG $\C'=(I',V,E,O')$ and another
$\C=(I,V,E,O)$ built from $\C'$ by just transforming some vertices without
predecessors into input vertices, and some others into output nodes so that
$I'\subset I$ and $O'\subset O$.
In contrast to the prior development above,
instead of adding/removing input/output vertices, here we do not change the
vertices of a CDAG but instead only change the labeling (tag) of some vertices
as inputs/outputs in the CDAG. So the CDAG remains the same, but some
input/output vertices are relabeled as standard computational vertices, or 
vice-versa.

\begin{theorem}[Input/Output (Un)Tagging -- RBW]
\label{thm:io-compar}
~\\
Let $C$ and $C'$ be two CDAGs of the same DAG $G=(V,E)$: $C=(I,V,E,O)$, $C'=(I\cup \deltaI,V,E,O\cup \deltaO)$. Then,
$\IO(C)$ can be bounded by a lower bound on $\IO(C')$ as follows (tagging):
\begin{equation}
\label{eq:tag}
\IO(C')-|\deltaI|-|\deltaO|\leq \IO(C)
\end{equation}
Reciprocally, $\IO(C')$ can be bounded by a lower bound on $\IO(C)$ as follows (untagging):
\begin{equation}
\label{eq:untag}\IO(C) \leq \IO(C')
\end{equation}
\end{theorem}
\comment{
\proof
Consider an optimal valid game $\cal P$ for $C$, of cost $\IO(C)$. We will build a valid game $\cal P'$ for $C'$, of cost no more than $\IO(C)+|\deltaI|+|\deltaO|$. This will prove that $\IO(C') \leq \IO(C)+|\deltaI|+|\deltaO|$. We build $\cal P'$ from $\cal P$ as follows: (1) for any input vertex $v\in \deltaI$, the first (and only) transition $R3$ involving $v$ in $\cal P$ is replaced in $\cal P'$ by a transition $R1$;
(2) for any output vertex $v\in \deltaO$,
the last transition $R3$ involving $v$ in $\cal P$ is complemented by an $R2$ transition;
(3) any other transition in $\cal P$ is reported as is in $\cal P'$.

Consider now an optimal valid game say $\cal P'$ for $C'$, of cost $\IO(C')$. We will build a valid game $\cal P$ for $C$, of cost no more than $\IO(C')$. This will prove that $\IO(C)\leq \IO(C')$. We build $\cal P$ from $\cal P'$ as follows:
(1) for any input vertex $v\in \deltaI$, the first transition $R1$ involving $v$ in $P'$ is replaced in $\cal P$ by a transition $R3$ followed by a transition $R2$;
(2) any other transition in $\cal P'$ is reported as is in $\cal P$.
\myendproof
}
}

Some algorithms will benefit from decomposing their CDAGs into
non-disjoint vertex sets. For instance, when we have computations that
are surrounded by an outer time loop, a common technique to derive
their lower bound is to decompose the CDAG, where vertices computed during
each outer loop iteration are placed in separate sub-CDAGs. In such
cases, when the vertices, $V$, computed in iteration $t$ are used as
inputs for iteration $t+1$, by placing $V$ in the
sub-DAGs corresponding to both iterations $t$ and $t+1$, we could
obtain a lower bound that is tighter by atleast a constant factor. 


%
\begin{theorem}[Non-disjoint Decomposition]
\label{thm:nondisjoint-decomp}
Consider an optimal game $\P$ of $C$ with $S$ red pebbles.
We let $Q_{L1}$ be the number of R1 transitions (loads) in $C$ associated to
a vertex of $C - [D_x + {x}]$.
We let $Q_{S1}$ be the number of R2 transitions (stores) in $C$ associated to
a vertex of $C-D_x$.
We let $Q_2$ be the number of R1 and R2 transitions (loads/stores) in
$C$
associated to a vertex in $D_x$.
We have that $IO_S(C)>= Q_{L1}+Q_2+Q_{S1}$.
\end{theorem}

\proof
The idea of the proof is to show that $Q_{L1}+Q_{S1}>=IO_{S+1}(C)$ and that
$Q2>=IO_S(C2)$.

Let us start with $Q_{L1}+Q_{S1}>=IO_{S+1}{C}$.
We consider the restriction of $C$ to the vertex of $C_1$. This is not a
valid game for $C_1$ yet, as the predecessors of any vertex in
$In(D_x)$ are
not the same in $C_1$ than in $C$. But we have one more red pebble that we
dedicate to stay on $x$. As soon as it is computed: we remove any R1
(loads) and R4 (delete) transitions associated to vertex $x$. This gives
a valid game for C1:
\begin{compactitem}
 \item as $C-D_x$ is a sub-graph of $C$ all transitions associated to a vertex
 of $C-[D_x+{x}]$ plus the transition R3 (compute) of x are valid (this
 part of the game has been unchanged).
  \item for a vertex in $D_x$ the only kept transitions are R3 / R2(compute /
  store) and is valid as all its predecessors are in $C-D_x$ which
  associated transitions are unchanged (apart from $x$ which keeps a red
  pebble as soon as it is computed).
  The cost of this valid game (with $S+1$ red pebbles) for C1 is
  $Q_{L1}+Q_{S1}$
  \end{compactitem}
  which proves the inequality.

  Let us now prove that $Q_{L2}+Q_{S2}>=IO_S(C2)$.
  We consider the restriction of $C$ to the vertex of $C_2=D_x$. This is a
  valid game for $C_2$ of cost $Q2$ which proves the second inequality.
\myendproof

\subsection{Min-Cut for I/O Complexity Lower Bound}
\label{sec:mincut}
In~\cite{LB-TR}, we developed an alternative lower bounding approach.
It was motivated from the observation that the \hk 2S-partitioning
approach does not account for the internal structure of a CDAG, but essentially
focuses only on the boundaries of the partitions. In contrast, the
min-cut based approach captures internal space requirements using the abstraction of wavefronts. 
This section describes the approach.

\noindent{\bf Definitions}: We first present needed definitions.
Given a graph $G=(\V,E)$, a cut is defined as any partition of the set
of vertices $\V$ into two parts $\Sset$ and $\Tset=\V-\Sset$. An $s-t$ cut is
defined with respect to two distinguished vertices $s$ and $t$ and is
any $(\Sset,\Tset)$ cut satisfying the requirement that $s \in \Sset$ and $t \in
\Tset$. Each cut defines a set of cut edges (the cut-set), i.e., the set of edges
$(u,v)$ where $u \in \Sset$ and $v \in \Tset$. The capacity of a cut is
defined as the sum of the weights of the cut edges.  The minimum cut
problem (or min-cut) is one of finding a cut that minimizes the
capacity of the cut. 
We define vertex $u$ as a cut vertex with respect to an $(\Sset,\Tset)$ cut,
as a vertex $u \in \Sset$ that has a cut edge incident on it.  A related
problem of interest for this paper is the \emph{vertex min-cut}
problem which is one of finding a cut that minimizes the number of cut
vertices.


We consider a convex cut $(\Sset_x,\Tset_x)$ associated to $x$ as follows: $\Sset_x$ includes $x\cup \Pred{x}$; $\Tset_x$ includes $\Succ{x}$;
in addition, $\Sset_x$ and $\Tset_x$ must be constructed such that there is no
edge from $\Tset_x$ to $\Sset_x$. With this, the sets $\Sset_x$ and $\Tset_x$
partition the graph $G$ into two convex partitions.  We define the wavefront 
induced by $(\Sset_x,\Tset_x)$ to be the set of vertices in
$\Sset_x$ that have at least one outgoing edge
to a vertex in $\Tset_x$.

\noindent{\bf Schedule Wavefront}: Consider a pebble game instance $\P$ that corresponds to some scheduling
(i.e., execution) 
of the vertices of the graph $G=(\V,E)$ that follows the rules R1--R4
of the Red-Blue-White pebble game (see Definition~\ref{def:rbwpg} in
Sec.~\ref{sec:rbwpg}).
We view this pebble game instance as a string that has recorded all
the transitions (applications of pebble game rules).  Given $\P$, we
define the \emph{wavefront} $W_\P(x)$ induced by some vertex $x \in \V$ at
the point when $x$ has just fired (i.e., a white pebble has just been
placed on $x$) as the union of $x$ and the set of vertices $u \in \V$
that have already fired and that have an outgoing edge
to a vertex $v \in \V$ that have not fired yet. Viewing the instance of the pebble game $\P$ as a string, $W_\P(x)$
is the set of vertices $x$ and those white-pebbled vertices to the
left of $x$ in the string associated with $\P$ that have an outgoing
edge in $G$ to not-white-pebbled vertices that occur to the right of
$x$ in $\P$.
With respect to a pebble game instance $\P$, the set $W_\P(x)$ defines
the memory requirements at the time-stamp just after $x$ has fired.

\noindent{\bf Correspondence with Graph Min-cut} Note that there is a one-to-one correspondence between the wavefront
$W_\P(x)$ induced by some vertex $x \in \V$ and the
$(\Sset_x,\Tset_x)$ partition of the graph $G$. For a valid convex
partition $(\Sset_x,\Tset_x)$ of $G$, we can construct a pebble game
instance $\P$ in which at the time-stamp when $x$ has just fired, the
subset of vertices of $\V$ that are white pebbled exactly corresponds
to $\Sset_x$; the set of fired (white-pebbled) nodes that have a
successor that is not white-pebbled constitute a wavefront $W_\P(x)$
associated with $x$.  Similarly, given wavefront $W_\P(x)$ associated
with $x$ in a pebble game instance $\P$, we can construct a valid
$(\Sset_x,\Tset_x)$ convex partition by placing all white pebbled
vertices in $\Sset_x$ and all the non-white-pebbled vertices in
$\Tset_x$.

A minimum cardinality wavefront induced by $x$, denoted
$W_G^{\textrm{min}}(x)$ is a vertex min-cut that results in an
$(\Sset_x,\Tset_x)$ partition of $G$ defined above.
We define $w^{\textrm{max}}_G$ as the maximum value over the size of
all possible minimum cardinality wavefronts associated with vertices,
i.e., define $w^{\textrm{max}}_G = \max_{x\in \V}
\left(\card{W_G^{\textrm{min}}(x)}\right)$.

\begin{lemma}
\label{lemma:wfnopart}
Let $C=(\emptyset,V,E,O)$ be a CDAG with no inputs. For any $x\in \V$, \hfill
$2\left(\card{W_G^{\textrm{min}}(x)}-\S\right)\leq \IO(C)$.\\
In particular, \hfill
$2\left(w^{\textrm{max}}_G-\S\right)\leq \IO(C)$.\\
\end{lemma}

\subsection{Parallel Red-Blue-White (P-RBW) Pebble Game}
\label{sec:parallel-rbw}
\begin{figure}[h!tb]
\ifdefined\isRR\else \vspace{-.3cm} \fi
\centering
\includegraphics[width=\ifdefined\isRR 0.6\fi\columnwidth]{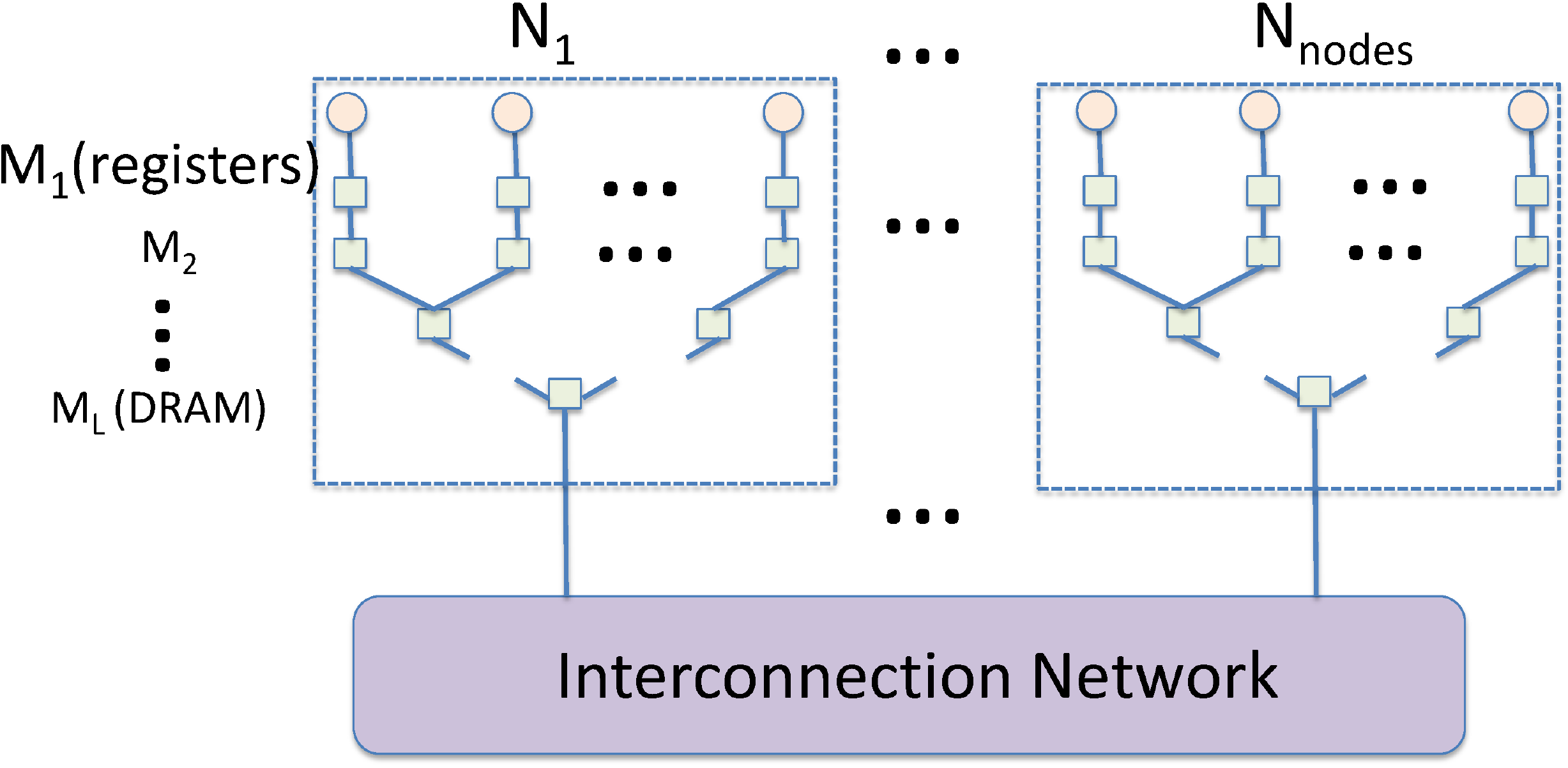}
\caption{\label{fig:arch}Distributed-memory system}
\ifdefined\isRR\else \vspace{-.3cm} \fi
\end{figure}

In this section, we extend the RBW pebble game to the parallel
environment.
P-RBW assumes that multiple nodes are
connected in a distributed environment, with each node containing
multiple cores. Further, the memory within each machine is organized in a hierarchical way. More formally, we have: (1) $N_L$ main memories (storage of level $L$) connected through ethernet; (2) $P$ processors, each of them having exactly $S_1$ registers (storage of level 1); (3) for each level ($1<l<L$), $N_l$ overall caches of size $S_l$ each; (4) one given cache of level $l$ has a unique (parent) cache of level $l+1$ to which it is connected. 
We consider that the bandwidth between a storage of level $l$ and its
children of level $l-1$ is shared between all its children. In other
words, the I/Os of the $P_l=P/N_l$ processors associated to a given
level $l$ storage instance are to be done sequentially. Note that
those $P/N_l$ processors have $S_{l-1}\times N_{l-1}/N_l$ storage
available at level $l-1$. Fig.~\ref{fig:arch} illustrates
this setup.

\begin{definition}[Parallel RBW (P-RBW) pebble game]
\label{def:prbwpg}
Let $\C=(I,V,E,O)$ be a CDAG. Given for each level $1\leq l\leq L,$ 
$N_l\times S_l$ number of red pebbles of different shades
$R^1_l,~R^2_l,\cdots,~R^{N_l}_{l}$, respectively, and unlimited blue
and white pebbles, with a blue pebble on each \textit{input} vertex, a
complete game is any sequence of steps using the following rules that
results in a final state with white pebbles on all vertices and blue
pebbles on all \textit{output} vertices:
\begin{compactitem}
\item[\textbf{R1 (Input)}] A level-$L$ pebble, $R^i_L$ can be placed
on any vertex that has a blue pebble; a white pebble is also placed
along with the shade of red pebble, unless the vertex already has a
white pebble on it.
\item[\textbf{R2 (Output)}] A blue pebble can be placed on any vertex that 
has a level-$L$ pebble on it.
\item[\textbf{R3 (Remote get)}] A level-$L$ pebble, $R^i_L$ can be placed
on any vertex that has another level-$L$ shade pebble $R^j_L$.
\item[\textbf{R4 (Move up)}] For $1\leq l<L$, a level-$l$ red pebble,
	$R^i_l$ can be placed on any vertex that has a level-$l+1$ pebble
	$R^j_{l+1}$ 
	where $R^i_l$ is in a cache that is a child of the cache that
        holds $R^j_{l+1}$.
\item[\textbf{R5 (Move down)}] For $1< l\leq L$, a level-$l$ red pebble,
	$R^j_l$ can be placed on any vertex that has a level-$l-1$ pebble
	$R^i_{l-1}$ where 
        $R^i_{l-1}$ is in a cache that is a child of the cache that 
        holds $R^j_{l}$.
\item[\textbf{R6 (Compute)}] If a vertex $v$ does not have a white
pebble and all its immediate predecessors have level-1 red pebbles on
them, then a level-1 red pebble $R^p_1$ along with a white pebble may be
placed on $v$; here $p$ is the index of the processor that 
comuptes vertex $v$.
\item[\textbf{R7 (Delete)}] Any shade of red pebble may be removed
from any vertex (reuse storage).
\end{compactitem}
\end{definition}

\section{I/O Lower Bound for parallel machines}
\label{sec:parallel-io}
In this section, we provide the necessary tools to analyze the
lower bounds for the parallel case. In particular, we consider two
distinct cases of data movement:
\begin{compactenum}
\item \emph{data movement along the memory
hierarchy within a processor}, which we call the \textbf{vertical data
movement}; 
\item \emph{data movement across processors}, which we call the
\textbf{horizontal data movement}.
\end{compactenum}

\subsection{I/O Lower Bound for Vertical Data Movement}
\label{sec:lb-vertical}
The hierarchical memory can enforce either the inclusion or exclusion
policy. In case of inclusive hierarchical memory, when a copy of a
value is present at a level-$l$, it is also maintained at all the
levels $l+1$ and higher. These values may or may not be consistent
with the values held at the lower levels. The exclusive cache, on the
other hand, does not guarantee that a value present in the cache at
level-$l$ will be available at the higher levels $\geq l+1$.
The following result is derived for the inclusive case. But, they also
hold true for the exclusive case, where the difference lies only in
the number of red pebbles that we consider in the corresponding two-level pebble game.

\begin{theorem}[Vertical I/O Cost]
\label{thm:vertical-io}
~\\
Let $\C=(I,V,E,O)$ be a CDAG. Consider any valid P-RBW game on $C$;
for this valid game, consider the level-$l$ storage $j$ with the
maximum number of $R5$ transitions placing a $R^j_l$ shade red pebble. The corresponding amount of move down transitions to this shade is at least $IO_1(C, S_{l-1}\times N_{l-1})/N_l$, where $IO_1(C,S)$ is the I/O lower bound of $C$ for a single processor with local memory of size $S$.
\end{theorem}
\proof
Consider a P-RBW game of $C$ that minimizes the overall amount of I/O between levels $k<l$ and level $l$ storage. 
This amount of I/O will be bounded by $IO_1(C, S_{l-1}\times N_{l-1})$. 
Consider one of the $N_l$ caches with the maximum amount of I/O. 
It will be bounded by $IO_1(C, S_{l-1}\times N_{l-1})/N_l$.
\myendproof

\begin{theorem}[Vertical I/O Cost]
~\\
\label{thm:vertical-io:2spart}
Let $\C=(I,V,E,O)$ be a CDAG. Consider any valid P-RBW game on $C$;
for this valid game, consider the level-$l$ storage $j$ with the
maximum number of $R5$ transitions placing a $R^j_l$ shade red pebble. The corresponding amount of move down transitions to this shade is at least\\ $\left[|V|/(U(C,2S_{l-1})\times N_l)-N_{l-1}/N_l\right]\times S_{l-1}\approx \frac{|V|\times S_{l-1}}{U(C,2S_{l-1})\times N_l}$ where $|V|$ is the total amount of work, $U(C,2S)$ is the largest 2S-partition of the CDAG $C$.
\end{theorem}
\proof
Consider a P-RBW game of $C$ that minimizes the overall amount of I/O between levels $k<l$ and level-$l$ storage. Consider the group of $P/N_l$ processors that do the more computation in this game. They do at least $|V|/N_l$ amount of work. Let us consider the partition of those $P/N_l$ processors into $N_{l-1}/N_l$ sets of $P/N_{l-1}$ processors that share the same level-$l-1$ storage unit. Each set of processor (that we denote $P^i$ with $0<i\leq N_{l-1}/N_l$) does at least $\alpha^i\times \frac{|V|}{N_l}$ amount of work where $\sum_i \alpha^i = 1$. We let $V^i$ be the subset of nodes of $C$ fired by $P^i$.

Let us denote $S_{l-1}$ by $S$ to simplify the notations. The goal is to show that each $P^i$ performs at least $\left[|V|^i/U(C,2S)-1\right]\times S$ I/O to its level-$l$ storage where $U(C,2S)$ is the largest 2S-partition (RBW pebble game) of CDAG $C$.
Consider an RBW game of $C$ with $S$ red pebbles. Consider the partitioning of the game into $C_1,\cdots,C_h$ used in the proof of Theorem~\ref{thm.hk} for RBW. We let $V_j^i$ be the set of vertices of $V^i$ fired in $C_j$ ($\bigcup V_j^i = V^i$; $V_j^i \cap V_j'^i=\emptyset$ for $j\neq j'$). With the usual reasoning we can prove that $|In(V_j^i)|\leq 2S$ and $|Out(V_j^i)|\leq 2S$ ie each $V_j^i$ is a 2S-partition of $C$.
Thus for each $j$, $|V_j^i|\leq U(C,2S)$.
Now from a valid P-RBW game, we can build a valid RBW game where the restriction to $V^i$ matches the P-RBW game. By construction, each $V_j^i$ is associated to at least $S$ I/O to level-$l$ storage in the P-RBW game. Thus the total amount of I/O for $P^i$ is at least $\left[|V^i|/|V_j^i|-1\right]\times S\geq \left[W^i/U(C,2S)-1\right]\times S$.

If we sum up the I/O of each set of processors with our level-$l$ storage unit associated to the $P/N_l$ processors that do the more computation in the game we get $\left[W/N_l-N_{l-1}/N_l\times U(C,2S)\right]\times S/U(C,2S) = \left[W/(U(C,2S).N_l)-N_{l-1}/N_l\right]\times S$ 
\myendproof

\subsection{I/O Lower Bound for Horizontal Data Movement}
\label{sec:lb-horizontal}
The following theorem extends the \S-partitioning technique to the
horizontal case.

\begin{theorem}[Horizontal I/O Cost]
~\\
\label{thm:parallel-spart}
Let $\C=(I,V,E,O)$ be a CDAG.
Consider any valid P-RBW game on $\C$; for this valid game, consider
the level-$L$ storage $i$ whose group of processors $P^i$ perform the maximum number of $R6$ (compute)
transitions. The corresponding amount of remote get transitions is
atleast $\left(\frac{\card{V}}{U(C,2S_L).P_i}-1\right)\times S_L$
\end{theorem}

\proof

We let $V^i$ be the subset of nodes of $\C$ fired by $P^i$. Let us
denote $S_L$ by $\S$ for simplicity.
Consider an RBW game of $C$ with $S$ red pebbles. Consider the
partitioning of the game into $C_1,\cdots,C_h$ used in the proof of
Theorem~\ref{thm.hk} for RBW. We let $V_j^i$ be the set of vertices of
$V^i$ fired in $C_j$ ($\bigcup V_j^i = V^i$; $V_j^i \cap
V_j'^i=\emptyset$ for $j\neq j'$). With the usual reasoning we can
prove that $|In(V_j^i)|\leq 2S$ and $|Out(V_j^i)|\leq 2S$ ie each
$V_j^i$ is a 2S-partition of $C$.
Thus for each $j$, $|V_j^i|\leq U(C,2S)$.
Now from a valid P-RBW game, we can build a valid RBW game where the
restriction to $V^i$ matches the P-RBW game. By construction, each
$V_j^i$ is associated to at least $S$ I/O operations in the
P-RBW game. Thus the total amount of I/O for $P^i$ is at least
$\left[|V^i|/|V_j^i|-1\right]\times S\geq
\left[W^i/U(C,2S)-1\right]\times S$.

Since the group $P^i$ performs maximum number of computations, $W^i
\ge W/P$. Hence, the total amount of remote get of processors $P^i$ is
atleast $\left((\card{V}/(U(C,2S_L).P_i))-1\right)\times S_L$
\myendproof

\section{Evaluation}
\def\LB{LB}
\def\UB{UB}
\def\nn{N_{nodes}}
\def\nc{N_{cores}}
\def\BW{\mathcal{B}}
A processor's machine balance is the ratio of the peak memory
bandwidth to the peak floating-point performance. Lower and upper
bound analysis of the algorithms can help us identify whether an
algorithm is bandwidth bound at different levels of memory hierarchy.

The lower bound results can be related to the architectural
machine balance as below:
Consider a multi-node/multi-core system with $P$ processors. 
Let $N_l$ be the total number of memory units available at level $l$.
Consider a memory unit at level $l$, $M_l^i$, that incurs the maximum communication.
$M_l^i$, is shared by the
processor set $P_l^i$, such that $\card{P_l^i} = P/N_l$.
Let
$\BW_l^i$ denote the total available memory bandwidth between 
$M_l^i$ and all its children at level $l-1$.

Let $\C=(I,V,E,O)$ be the CDAG of the algorithm being analyzed and
$\C^{l,i} \subset C$ be the sub-CDAG executed by the processors
$P_l^i$.
The time taken for execution of $\C$ is given by $$T \ge
max(T_l^i, T_{comp})$$
where, $T_l^i$ denotes the communication time at $M_l^i$
and $T_{comp}$ denotes the computation time for $\C$.

For the algorithm to be not bound by memory at level $l$,
\begin{equation}\label{eq:mem-bound}
	T_l^i \le T_{comp}
\end{equation}
Let $\IO_l^i$ denote the amount of data transferred between $M_l^i$
and all its children at level $l-1$ for the execution of a
$C_l^i$. Then,
\begin{equation}\label{eq:comm-time}
	T_l^i  =  \frac{\IO_l^i}{\BW_l^i} \ge
	\frac{\LB_l^i}{\BW_l^i}
\end{equation}
where, $\LB_l^i$ denotes the lower bound on the amount of data
transfer at memory unit $M_l^i$ for any valid execution of
$\C_l^i$.
The computation time of $\C$ is given by
($F$ below indicates FLOPs)
\begin{equation}\label{eq:comp-time}
	T_{comp} \ge \frac{\card{V}}{P}\times \frac{1}{F}
\end{equation}
From Equations (\ref{eq:mem-bound}),~(\ref{eq:comm-time})
and~(\ref{eq:comp-time}), we
have,
$$
\frac{\LB_l^i}{\BW_l^i} \leq \frac{\card{V}}{P}\times \frac{1}{F}
\quad \mbox{ or } \quad
\frac{\LB_l^i}{\card{V}} \leq \frac{\BW_l^i}{P}\times \frac{1}{F}
$$
As $P = \card{P_l^i}\times N_l^i$,
\begin{eqnarray}
\label{eq:mc-bal-lb}
	\frac{\LB_l^i\times N_l^i}{\card{V}} & \leq &
	\frac{\BW_l^i}{\card{P_l^i}\times {F}}
\end{eqnarray}
The term at the right-hand side of equation~\ref{eq:mc-bal-lb} is the
machine balance value for the machine.

Hence, any algorithm that fails to satisfy the condition~\ref{eq:mc-bal-lb},
will be bandwidth bound at level $l$ irrespective of any
optimizations we do to the code.

Through similar argument, given that $\UB_l^i$ is the upper bound on
the minimum amount of data transfer required by the algorithm at
memory unit $M_l^i$,
we can show that if the algorithm is communication bound, then it
definitely satisfies the condition,
\begin{equation}\label{eq:mc-bal-ub}
	\frac{\UB_l^i\times N_l^i}{\card{V}} \geq
	\frac{\BW_l^i}{\card{P_l^i}\times {F}}
\end{equation}
Hence, if an algorithm fails to satisfy condition~\ref{eq:mc-bal-ub},
we can safely conclude that there is atleast one execution
order of $\C$ that is not constrained by the memory bandwidth at level
$l$.

In particular, we were interested in understanding the memory
bandwidth requirements (1) between the main memory and L2 cache within
each processor, and, (2) between different processors for various
algorithms. For simplicity, we assume that the L2 cache is shared by
all the cores within a node, which is common in practice. Our claim
is that the vertical data movement between the main memory and L2
cache will be the major bottleneck in the future machines, compared to
the inter-node data movement. We show that this is true for various
algorithms by comparing the lower bound for vertical data movement and
the upper bound for horizontal data movement against the machine
balance values of different machines.

Considering the particular case of data movement between L2 cache and
the main memory, equation~\ref{eq:mc-bal-lb} becomes,
\begin{eqnarray}\label{eq:mc-bal-vert}
	\frac{\LB_{vert}\times N_{nodes}}{\card{V}} & \leq &
	\frac{\BW_{vert}}{N_{cores}\times {F}}
\end{eqnarray}
where, $\LB_{vert}$ is the vertical data movement lower bound,
$\BW_{vert}$ is the total bandwidth between DRAM and L2 cache, $N_{nodes}$ represents the number of nodes in the system, and
$N_{cores}$ represents the number of cores within each node.
Similarly, considering the inter-node communication,
equation~\ref{eq:mc-bal-ub} becomes,
\begin{eqnarray}\label{eq:mc-bal-horiz}
	\frac{\UB_{horiz}\times N_{nodes}}{\card{V}} \geq
\frac{\BW_{horiz}}{N_{cores}\times {F}}
\end{eqnarray}
where, $\UB_{horiz}$ and $\BW_{horiz}$ represent the upper bound on
the horizontal data movement cost and inter-processor communication
bandwidth, respectively.

Specifications for some of the powerful computing systems are
shown in table~\ref{tb:mc-bal-list}. We plan to use this list to compare
various algorithms in the following sections to determine their memory
requirement constraints.

{\small \begin{table}[h!tb]
\centering
\caption{\label{tb:mc-bal-list}Specifications of various computing systems}
\begin{tabular}{l || p{.75cm} | p{.75cm} | p{.75cm} | p{1.5cm} | p{1.5cm}}
	\hline
	\textbf{Machine} & {$\mathbf\nn$} & \textbf{Mem. (GB)} &
  \textbf{L2/L3 cache (MB)} & \textbf{Vertical
		      balance (words/FLOP)} & \textbf{Horiz.
balance (words/FLOP)}\\
	\hline
IBM BG/Q & 2048 & 16 & 32 & 0.052 & 0.049\\
Cray XT5  & 9408 & 16 & 6 & 0.0256 & 0.058 \\
	\hline
\end{tabular}
\end{table}
}

Before we present the results for various numerical algorithms, we
provide a brief introduction to the type of problem solved by these
numerical solvers, in the following section.

\subsection{Brief introduction on discretization}
\label{ap:cg-intro}
\begin{figure}[h!tb]
\ifdefined\isRR\else \vspace{-.3cm} \fi
\centering\includegraphics[width=\ifdefined\isRR 0.6\else .85\fi\columnwidth]{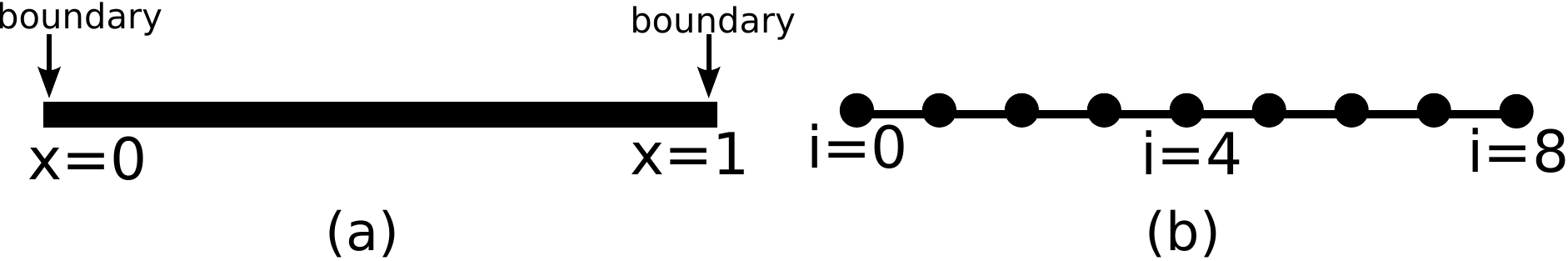}
\caption{\label{fig:1d-prob}One-dimensional heat flow
problem}
\ifdefined\isRR\else \vspace{-.3cm} \fi
\end{figure}

Many real world problems involve solving
partial differential equations (PDEs). As an example, consider the
heat flow on a long thin bar of unit length, of uniform material and
insulated, so that the heat can enter and exit only at the boundaries
(refer Fig.~\ref{fig:1d-prob}(a)).  Let $u(x,t)$ represent the temperature at
position $0\le x\le 1$, and time $t\ge 0$. The objective is to
determine the change in temperature over time ($u(x,t)$). The
governing \emph{heat equation} that describes this distribution of
heat is given by the PDE:
$$\frac{du(x,t)}{dt} = \alpha \times \frac{d^2u(x,t)}{dx^2}$$
where, $\alpha$ is the thermal diffusivity of the bar. (For 
mathematical treatment, it is sufficient to consider $\alpha=1$).

Since the problem is continuous, to numerically solve the heat equation, it needs
to be \emph{discretized} (through \emph{finite difference}
approximation) to reduce it to a finite problem.
In the discretized problem, the values of $u(x,t)$ are only computed at discrete points at
regular intervals of the bar, called the \emph{computational grid} or
\emph{mesh}. The state variables at these grid points are given by $u(x(i),t(m))$, where $x(i)=i\times
h,~0\le i\le n+1=1/h$ and $t(m) = m\times k$; $h$ and $k$ are the
\emph{grid
spacing} and \emph{timestep}, respectively. 
Fig.~\ref{fig:1d-prob}(b)
shows an example grid obtained by discretizing the one-dimensional bar.

The governing equation, after discretization, yields the following
equation at grid point $i$ and timestamp $m+1$.
\newcommand{\z}{a}
\newcommand{\temp}[2]{U(#1,#2)}
\begin{eqnarray*}
\frac{-\z}{2}\times \temp{i-1}{m+1} + (1+\z) \times \temp{i}{m+1} -
\frac{\z}{2}\times \temp{i+1}{m+1} = \\
\frac{\z}{2}\times \temp{i-1}{m} +
(1-\z)\times \temp{i}{m} + \frac{\z}{2} \times \temp{i+1}{m}
\end{eqnarray*}
where, $\temp{p}{q}=u(x(p),t(q))$ and $\z=k/h^2$.
Hence, the solution to the problem involves solving a linear system of
$n-1$ equations at each timestamp till convergence.
Each timestamp
$m+1$ is dependant on values of the previous timestamp $m$.

This linear system can be
represented in tridiagonal matrix form as follows:
\begin{equation}
\label{eq:1d-matrix}
\left(
\begin{smallmatrix}
1+\z & -\frac{\z}{2} &  &  &  &  &  \\
-\frac{\z}{2} & 1+\z & -\frac{\z}{2} &  & & &  \\
 & -\frac{\z}{2} & 1+\z & -\frac{\z}{2} & & & \\
& & \ddots & \ddots & \ddots & & &\\
& & & -\frac{\z}{2} & 1+\z & -\frac{\z}{2} &\\
& & & & -\frac{\z}{2} & 1+\z\\
\end{smallmatrix}
\right)
\times
\left(
\begin{smallmatrix}
\temp{1}{m+1} \\
\temp{2}{m+1} \\
\temp{3}{m+1} \\
\vdots \\
\temp{n-1}{m+1} \\
\temp{n}{m+1} \\
\end{smallmatrix}
\right)
=
\left(
\begin{smallmatrix}
b(1,m) \\
b(2,m) \\
b(3,m) \\
\vdots \\
b(n-1,m) \\
b(n,m)
\end{smallmatrix}
\right)
\end{equation}
where, $b(i,m)$ represents the right-hand side of the $i$-th equation at
timestamp $m+1$.
Solving this linear system of the form $Ax=b$ for vector $x$ provides
the solution to the original problem.
In general, for a $d$-dimensional problem, the coefficient matrix is
of size $n^d$-by-$n^d$, while the vectors are of size $n^d$. In
practice, the elements of the matrix are not explicitly stored.
Instead, their values are directly embedded in the program as
constants thus eliminating the space requirement and the associated
I/O cost for the matrix.

In practice, the prohibitive problem size prevents direct
solution of (\ref{eq:1d-matrix}). Hence, various
iterative methods were developed to efficiently
solve such large linear systems. The following section derives the vertical
and horizontal data movement bounds for some of these iterative
linear system solvers using the results from Sections
\ref{sec:parallel-game} and \ref{sec:parallel-io} and compares it
against the machine balance values.

\subsection{Conjugate Gradient (CG)}
\label{sec:cg-overview}
The Conjugate Gradient method~\cite{hestenes1952methods} is suitable
for solving symmetric positive-definite linear systems.
CG maintains 3 vectors at each timestep - the
approximate solution $x$, its residual $r=Ax-b$, and a search
direction $p$. At each step, $x$ is improved by searching for a better
solution in the direction $p$.

Each iteration of CG involves one sparse matrix-vector product, three vector
updates, and three vector dot-products. The complete pseudocode is shown
in Fig.~\ref{fig:cg-algo}.

\begin{figure}[h!tb]
\ifdefined\isRR\else \vspace{-.3cm} \fi
\centering\begin{minipage}{.9\columnwidth}
{\small
\begin{lstlisting}[basicstyle=\small,frame=single,mathescape,numbers=left,escapechar=\#]
function CG
  $x$ is the initial guess
  $r \leftarrow b - Ax$
  $p \leftarrow r$
  #\DO#
    $v\leftarrow Ap$#\hfill#//SpMV #\label{ln:cg-spmv}#
    $a \leftarrow \inprod{r,r}/\inprod{p,v}$#\hfill#//Dot-products #\label{ln:cg-dota}#
    $x \leftarrow x + ap$#\hfill#//saxpy #\label{ln:cg-p}#
    $r_{new}\leftarrow r - av$#\hfill#//saxpy #\label{ln:cg-v}#
    $g \leftarrow \inprod{r_{new},r_{new}}/\inprod{r,r}$#\hfill#//Dot-products #\label{ln:cg-dotg}#
    $p\leftarrow r_{new} + gp$#\hfill#//saxpy #\label{ln:cg-r}#
    $r\leftarrow r_{new}$
  #\UNTIL# $\left(\inprod{r_{new},r_{new}}~is~small~enough\right)$
end function
\end{lstlisting}
}
\end{minipage}
\ifdefined\isRR\else \vspace{-.2cm} \fi
\caption{\label{fig:cg-algo}Conjugate Gradient method}
\ifdefined\isRR\else \vspace{-.3cm} \fi
\end{figure}

\subsubsection{Vertical data movement cost}
We provide the lower bound for the amount of data movement between
different levels of hierarchy.

\begin{theorem}[Min-cut based I/O lower bound for CG]
For a $d$-dimensional grid of size $n^d$, the minimum I/O cost to solve the
linear system using CG, $Q$, satisfies $Q\geq 6n^dT/P$, when
$n\gg\S$;
where, $T$ represents the number of outer loop iterations.
\end{theorem}

\proof
Consider the vertex $\upsilon_x$, corresponding to the
scalar $a$ at line~\ref{ln:cg-dota}. The $2n^d$ predecessor vertices
of $\upsilon_x$, corresponding to vectors $p$ and $v$, have disjoint paths to
the $\Succ{\upsilon_x}$ (due to computations in lines~\ref{ln:cg-p}
and~\ref{ln:cg-v}, respectively). This gives us a wavefront of size
$\card{W_G^{min}(\upsilon_x)} = 2n^d$. Similarly, considering the 
vertex, $\upsilon_y$, corresponding to the scalar $g$, at
line~\ref{ln:cg-dotg}, we obtain a wavefront of size
$\card{W_G^{min}(\upsilon_y)} = n^d$, due to the disjoint paths from
the predecessors $r_{new}$ to $\Succ{x}$ (due to the computation at
line~\ref{ln:cg-r}).

Recursively applying theorem~\ref{thm:nondisjoint-decomp} on the
complete CDAG, $\C$, provides us $T$ sub-CDAGs, $\C_1,~\C_2,\dots,\C_T$, corresponding to each
outer loop iteration. (Vertices of vector $p$ due to
line~\ref{ln:cg-r} are
shared between neighboring sub-CDAGs). Further, non-disjointly sub-dividing each of
these sub-CDAGs, $\C_i$, into $\C_{i_{|x}} and \C_{i_{|y}}$ (vertices of
vector $r_{new}$ from line~\ref{ln:cg-v} are shared between
$\C_{i_{|x}}
and \C_{i_{|y}}$), to decompose the effects of wavefronts
$W_G^{min}(\upsilon_x)$ and $W_G^{min}(\upsilon_y)$, we obtain a lower
bound of,
\begin{eqnarray*}
\Q & \geq & T\times (2(2n^d-S)) + T\times(2(n^d-S)) \\
	  & = & T\times (2(3n^d-2S))
\end{eqnarray*}
which tends to $6n^dT$ as $n$ becomes $\gg S$. Finally, application of
theorem~\ref{thm:vertical-io} provides a lower bound of $6n^dT/P$ for
the parallel case.
\myendproof

\subsubsection{Horizontal data movement cost}
\label{sec:cg-ub}
Let us assume that the input grid is block partitioned among the
processors. Hence, each processor holds the input data corresponding
to its local grid points and computes the data needed by those grid
points. Let $B = n/N_{nodes}^{1/d}$ be the size of the block along each dimension.
Computating the sparse matrix-vector product, at line~\ref{ln:cg-spmv}
in Fig.~\ref{fig:cg-algo}, majorly contributes to the communication cost.
This involves getting the values of the ghost cells from the
neighboring processors. This value is given by $(B+2)^d-B^d$.
If $\Q$ is the minimum I/O cost for executing CG, then,
\begin{eqnarray*}
\Q & \le & ((B+2)^d-B^d)\times T \\
	& = & (B^d + {d \choose 1}B^{d-1}2^1 + {d \choose 2}B^{d-2}2^2
	\\
	& & + \cdots + {d \choose d-1}B^1 2^{d-1} + {d \choose d}B^0 2^d
	B^d)\times T \\
	& = & O(2dB^{d-1}T)
\end{eqnarray*}

\comment{
\begin{theorem}[I/O Upper bound for CG]
Conjugate Gradient method (Fig.~\ref{fig:cg-algo}) can solve a
$d$-dimensional
linear system of $n^d$ grid points with an asymptotic I/O cost of
$9n^dT/P$,
where, $T$ represents the number of outer loop iterations.
\end{theorem}

\proof
To prove the upper bound, we directly estimate the I/O cost needed for
executing the algorithm in Fig.~\ref{fig:cg-algo}.
The I/O cost in the
CG algorithm arises due to the loads and stores of the vectors
$p,~r,~v,~x$ and $r_{new}$, each of sizes $n^d/P$ per processor.
Lines~\ref{ln:cg-dota},~\ref{ln:cg-p},~\ref{ln:cg-v}
and~\ref{ln:cg-r} contribute to the I/O cost of
$3n^d/P,~2n^d/P,~3n^d/P$ and $2n^d/P$, respectively, due to the
vector loads and stores per outer loop iteration. This leads to an asymptotic I/O cost of
$9n^dT/P$.
\myendproof
}

\subsubsection{Analysis}
Equations~\ref{eq:mc-bal-vert} and~\ref{eq:mc-bal-horiz} provided us
conditions to determine the vertical and horizontal memory constraints
of the algorithms. We will use them to show that the running time of
CG is mainly constrained by the vertical data movement.

Consider a 3D-grid ($d=3$), with $n=1000$. The total operation count
(FLOP) is $20n^3T$. The I/O lower bound per node is
given by $\displaystyle\frac{6n^3T}{P}\times \nc = \frac{6n^3T}{\nn}$.
Hence,
$$
\frac{LB_{vert}\times \nn}{\card{V}}  = 
	\frac{\left(6n^3T/\nn\right) \times \nn}{20n^3T} 
	= \frac{6}{20} = 0.3
$$
This value is higher than the machine balance value of any machine
(refer table~\ref{tb:mc-bal-list}), leaving
condition~\ref{eq:mc-bal-vert} unsatisfied. This shows that CG will be
unavoidably bandwidth bound along the vertical direction for the
problems that cannot fit into the cache. The only way to improve the
performance would be to increase the main memory bandwidth.

On the other hand, let us consider the horizontal data movement cost.
\begin{eqnarray*}
& &	\frac{\UB_{horiz}\times \nn}{\card{V}} =
	\frac{6B^2T\times \nn}{20n^3T} \\
	& = & \frac{6\left(n/\nn^{(1/3)}\right)^2\nn}{20n^3} 
	= \frac{6\nn^{(1/3)}}{20n}
\end{eqnarray*}
\comment{
\begin{eqnarray*}
& &	\frac{\UB_{horiz}\times \nn}{\card{V}} =
	\frac{6B^2T\times \nn}{20n^3T} \\
<<<<<<< .mine
	& = & \frac{6\left(n/\nn^{(1/3)}\right)^2\nn}{20n^3} 
	= \frac{6\nn^{(1/3)}}{20n}
=======
	 =  \frac{6\left(n/\nn^{(1/3)}\right)^2\nn}{20n^3} & = & \frac{6\nn^{(1/3)}}{20n}
>>>>>>> .r20590
\end{eqnarray*}
}
This value easily falls below the machine balance values of various
machines, indicating that the inter-node communication is not a
bottleneck for the execution of CG algorithm.

\subsection{Generalized Minimal Residual (GMRES)}
The Generalized Minimum Residual (GMRES)~\cite{saad1986gmres} method
is designed to solve non-symmetric linear systems. The most popular
form of GMRES is based on the modified Gram-Schmidt procedure. The
method arrives at the solution after $m$ iterations, where $m$
represents the dimension of a linear subspace--called Krylov 
subspace--formed
by an orthogonal basis. At each step, $x$ is improved by searching for a
better solution along the direction of one of these basis vectors.

Each outer loop iteration $i$ of GMRES involves one sparse matrix-vector
product, $(i+1)$ vector dot-products and $i$ vector updates. The complete
pseudocode is shown in Fig.~\ref{fig:gmres-algo}.

\begin{figure}[h!tb]
\ifdefined\isRR\else \vspace{-.3cm} \fi
\centering\begin{minipage}{.9\columnwidth}
{\small
\begin{lstlisting}[basicstyle=\small,frame=single,mathescape,numbers=left,escapechar=\#]
function GMRES
  $x_0$ is the initial guess
  $r_0 \leftarrow b - Ax_0$
  $v_0 \leftarrow r_0/\norm{r_0}$
  #\DO#
    $w \leftarrow Av_i$#\label{ln:gmres-spmv}\hfill#//SpMV
    #\FOR# $j = 0,1,\dots,i$ #\DO#
      $h_{j,i} \leftarrow \inprod{w,v_j}$#\hfill#//Dot-product#\label{ln:gmres-dotprod}#
    #\ENDDO#
    $v'_{i+1} \leftarrow w - \sum_{j=1}^i h_{j,i}v_j$#\hfill#//saxpy's#\label{ln:gmres-x-desc}#
    $h_{i+1,i} \leftarrow \norm{v'_{i+1}}$#\hfill#//Dot-product#\label{ln:gmres-norm}#
    $v_{i+1} \leftarrow v'_{i+1}/h_{i+1,i}$ #\label{ln:gmres-vi1}#
    Apply Givens rotations to $h_{:,i}$
  #\UNTIL# (convergence)
  $y \leftarrow \argmin{\norm{Hy - \norm{r_0}e_1}}$
  $x \leftarrow x_0 + Vy$
end function
\end{lstlisting}
}
\end{minipage}
\ifdefined\isRR\else \vspace{-.2cm} \fi
\caption{\label{fig:gmres-algo}Basic GMRES}
\ifdefined\isRR\else \vspace{-.3cm} \fi
\end{figure}

\subsubsection{Vertical data movement cost}
\begin{theorem}[Min-cut lower bound for GMRES]
For a $d$-dimensional grid of size $n^d$, the minimum I/O cost to solve the
linear system using GMRES, $Q$, satisfies $Q\geq 6n^dm/P$, when
$n\gg\S$;
where, $m$ represents the number of outer loop interations.
\end{theorem}

\proof
Consider the vertex $\upsilon_x$ corresponding to the result of the inner product at
iteration $j=i$ at line~\ref{ln:gmres-dotprod}. This is a reduction
operation with the predecessors of $\upsilon_x$ being the vertices of vectors $w$
and $v_i$ of size $n^d$. All of these $2n^d$ predecessor vertices (of vectors
$w$ and $v_i$) have a disjoint path to the successor of vertex
$\upsilon_x$ due to
the computation at line~\ref{ln:gmres-x-desc}, leading to a wavefront of
size $\card{W_G^{min}(\upsilon_x)} = 2n^d$. By similar argument, by taking
vertex $\upsilon_y$ as the result of the
computation at line~\ref{ln:gmres-norm}, we obtain wavefront of size
$\card{W_G^{min}(\upsilon_y)} = n^d$ due to the vertices of vector $v'_{i+1}$.

Application of theorem~\ref{thm:nondisjoint-decomp} recursively allows us
to non-disjointly decompose
the vertices of each iteration of loop-$i$ into $m$ sub-DAGs,
$\C_1,~\C_2,\dots,\C_m$, with vertices
of vector $v_{i+1}$ (computed at line~\ref{ln:gmres-vi1}) being shared by
sub-DAGs $\C_i$ and $\C_{i+1}$. Each of these sub-DAGs, $\C_i$ can be further
non-disjointly partitioned into $\C_{i_{|x}} and \C_{i_{|y}}$, to decompose the effects of wavefronts
$W_G^{min}(\upsilon_x)$ and
$W_G^{min}(\upsilon_y)$ into separate sub-DAGs (with vertices of $v'_{i+1}$ from
	line~\ref{ln:gmres-x-desc} being shared between sub-CDAGs
$\C_{i_{|x}} and \C_{i_{|y}}$). This gives
us $m$ sub-DAGs, each with wavefront of size $2n^d$, and $m$ sub-DAGs with
wavefront of size $n^d$.

Applying Lemma~\ref{lemma:wfnopart} on these sub-DAGs gives us a lower
bound of,
\begin{eqnarray*}
\Q & \geq & m\times (2(2n^d-S)) + m\times(2(n^d-S)) \\
	& = & m\times (2(3n^d-S)) 
\end{eqnarray*}
which tends to $6n^dm$ as $n$ grows. Finally, application of
theorem~\ref{thm:vertical-io} provides a lower bound of $6n^dm/P$ for the
parallel case.
\myendproof

\comment{
\begin{theorem}[I/O upper bound for GMRES]
GMRES (Fig.~\ref{fig:gmres-algo}) can solve a
$d$-dimensional
linear system of $n^d$ grid points with an asymptotic I/O cost of $n^dm^2/P$.
\end{theorem}

\proof
To prove the upper bounds, we directly estimate the I/O cost required for executing the algorithm in
Fig.~\ref{fig:gmres-algo}. The reduction
operations in every iteration of the outer loop prevent any possible
tiling of the code for data reuse.

We do not count the cost of loading the input matrix $A$, since the
values of matrix are generally implicitly represented in
practice and hence doesn't need actual storage.
The inner $j$ loop is executed $i$ times per outer loop iteration; and
the $n^d/P$ input elements of vector $v_j$ need to be loaded
by each processor during each iteration. This amounts to the I/O cost of
$\sum\limits_{i=1}^m (i\times n^d/P) \approx
\frac{n^dm^2}{2P}$. Similarly, the I/O cost due to
line~\ref{ln:gmres-x-desc} is $\frac{n^dm^2}{2P}$. Loads due to the
elements of vector $v'_{i+1}$ in lines~\ref{ln:gmres-norm}
and~\ref{ln:gmres-vi1} leads to an I/O 
of $\frac{n^dm}{P}$ each.
This provides an I/O cost of $\displaystyle\frac{n^dm^2}{P} + \frac{n^dm}{P} =
O\left(\frac{n^dm^2}{P}\right)$.
\myendproof
}

\subsubsection{Horizontal data movement cost}
The horizontal data movement trend for GMRES is similar to CG
(Sec.~\ref{sec:cg-ub}). Hence,
upon applying similar analysis on the GMRES algorithm, we obtain an
upper bound of $$\Q = O(2dB^{d-1}m)$$
where, $B$ is the block size along each dimension.

\subsubsection{Analysis}
Consider a 3D-grid ($d=3$), with $n=1000$. The total number of
operations for GMRES is $20n^3m + n^3m^2$.
The vertical data movement cost per FLOP,
\begin{eqnarray*}
	\frac{LB_{vert}\times \nn}{\card{V}} & = & \frac{6}{m+20}
\end{eqnarray*}
For smaller values of $m$, this value stays higher than the machine
balance value for current systems. But as $m$ gets higher, the
computational time begins to dominate the vertical data movement cost.

The Horizontal data movement cost per FLOP is given by,
\begin{eqnarray*}
	\frac{\UB_{horiz}\times \nn}{\card{V}} & = &
	\frac{6\nn^{(1/3)}}{nm}
\end{eqnarray*}
This value is orders of magnitude smaller than the machine balance
values of current systems, showing that the algorithm is not
inter-node bandwidth bound.

On the other hand, for the vertical data movement, as the lower and
upper bounds do not match, it is not possible to draw a decisive
conclusion unless the rate of convergence value, $m$ is known for the
problem.

\subsection{Jacobi Method}
Jacobi's method involves stencil computations, that begins with an initial guess for the unknown vector
$x$ and iteratively replaces the current approximate solution at each grid 
point by a
weighted average of its nearest neighbors on the grid. Hence, the
information at one grid point can only propogate to its adjacent grid
points in one iteration. Thus, it takes atleast $n$ steps to propogate
the information throughout the grid and reach to the solution.

\subsubsection{Vertical data movement cost}
In this section, we derive the I/O lower bound for Jacobi computation
on a $d$-dimensional grid. We provide the proof for a 2D-grid below,
which can be generalized to a grid of $d$-dimensions as shown later.

\begin{theorem}[I/O lower bound for Jacobi]
\label{thm:lb-jacobi}
For the $9$-points Jacobi of size $n\times n$ with $T-1$ time
steps, the minimum I/O cost, $\Q$, satisfies $\Q \geq
\frac{N^2T}{4P\sqrt{2\S}}$.
\end{theorem}

\proof
%
The CDAG of Jacobi computation has the property that all inputs can
reach all outputs through vertex-disjoint paths. These vertex-disjoint
paths will be called \emph{lines}, for simplicity.
Let $F(d)$ denote a monotonically increasing function 
such that for any two vertices $u$ and $v$ on the same line
that are atleast $d$ apart, $F(d)$ has the following properties:
(1) none of these $F(d)$ vertices belong to the same line; (2) Each of
these vertices belongs to a path connecting $u$ and $v$.
In \cite[Theorem 5.1]{hong.81.stoc}, \hk show that the serial I/O lower
bound, $Q_s$, for the CDAG with the above mentioned properties can be bounded
by $Q_s \ge L/(2.(F^{-1}(2\S)+1))$, where $L$ is the total number of
vertices on the lines.
From the structure of the CDAG for 2D-Jacobi computation, it can be
seen that $F^{-1}(2S)=2\sqrt{2\S}-1$.
Hence, we have, $Q_s\ge n^2T/4\sqrt{2\S}$. Finally, from Theorem
\ref{thm:vertical-io}, we have the parallel I/O cost, $Q\ge
n^2T/4P\sqrt{2\S}$.
\myendproof

With the similar reasoning, the I/O lower bound can be extended to
higher dimensions, leading to the I/O cost of $Q\ge
n^dT/4.P.(2\S)^{1/d}$, for a $d$-dimensional grid.

It could be seen that this lower bound is tight as 
the tiled stencil computation algorithm has the I/O cost that matches
this bound.

\subsubsection{Horizontal data movement cost}
The horizontal data movement cost is due to the communication of the
ghost cells. This amounts to the I/O cost of $4BT$, where $B$ is the
block size along each dimension.

\subsubsection{Analysis}
%
From (\ref{eq:mc-bal-lb}) and Theorem \ref{thm:vertical-io:2spart},
and since the lower bound derived in Theorem \ref{thm:lb-jacobi} is
tight, for the computation to be not bandwidth-bound along the vertical
direction, the following relation has to be satisfied:
\begin{eqnarray*}
\frac{\BW^i_l}{\card{P^i_l}\times {F}} & \ge &
	\frac{\LB^i_l\times N^i_l}{\card{V}}\\
	& = & \frac{(\card{V}.\S_{l-1}/U(C,2S_{l-1}).N_l^i) \times
N_l^i}{\card{V}}\\
& = & \frac{\S_{l-1}}{U(C,2S_{l-1})}
\end{eqnarray*}

From Theorem \ref{thm:lb-jacobi}, for a $d$-dimensional Jacobi, $U(C,2S_{l-1}) =
4S_{l-1}(2S_{l-1})^{1/d}$. Hence,
\begin{equation*}
\frac{\BW^i_l}{\card{P^i_l}\times {F}}  \ge 
	\frac{1}{4(2S_{l-1})^{1/d}}
\end{equation*}
From Table \ref{tb:mc-bal-list}, for IBM BG/Q, the vertical machine
balance parameter for the data movement between main memory and L2
cache is 0.052. Hence,
${1}/{4(2S_{2})^{1/d}} \le 0.052$
or, $d\le 0.21\log(2\S_{2})$. Substituting the value of $S_2=4$
MWords, we get, $d \le 4.83$.

By following similar reasoning and considering the machine parameters
for the data movement between L2 and L1 caches, for the computation to
be not bandwidth-bound, $d\le96$.

This shows that the data movement between main memory and L2 cache is
critical for the preformance and the algorithm is bandwidth bound only for
higher dimensional stencils of dimension $d\ge5$, which are not common in practice.

\section{Related Work}
\label{sec:related}


\hk provided the first characterization of the I/O complexity problem
using the red/blue pebble game and the equivalence to 2S-partitioning
of CDAGs \cite{hong.81.stoc}.  Their 2S-partitioning approach uses
dominators of incoming edges to partitions but does not account for
the internal structure of partitions. In this paper, in addition to using
the 2s-partitioning technique, we also use an
alternate lower bound approach that models the internal structure of
CDAGs, and uses graph mincut as the basis.
In addition, \hk's original
model does not lend itself easily to development of effective lower bounds for a
CDAG from bounds for component sub-graphs. With a change of the pebble game
model to the RBW game, we were able to use CDAG decomposition to develop
tight composite lower bounds for inhomogeneous CDAGs. 

Several works followed \hk's work on I/O complexity in deriving lower
bounds on data accesses
\cite{aggarwal.ca.88,AACS87,toledo.jpdc,bilardi2000,bilardi2001characterization,savage.cc.95,savage.book,ranjan11.fft,ranjan12.rpyr,valiant.jcss.11,DemmelGHL12,BDHS11,BDHS11a,Demmel2013TR,apsp_ipdps2013,savage.options}.
Aggarwal et al. provided several lower bounds for sorting algorithms
\cite{aggarwal.ca.88}.  Savage \cite{savage.cc.95,savage.book}
developed the notion of $S$-span to derive Hong-Kung style lower
bounds and that model has been used in several works
\cite{ranjan11.fft,ranjan12.rpyr,savage.options}.
Irony et al.~\cite{toledo.jpdc} provided a new proof of the Hong-Kung
result on I/O complexity of matrix multiplication and developed lower
bounds on communication for sequential and parallel matrix
multiplication.
More recently, Demmel et al. have developed lower bounds as well as
optimal algorithms for several linear algebra computations including
QR and LU decomposition and all-pairs shortest paths problem
\cite{BDHS11,BDHS11a,DemmelGHL12,apsp_ipdps2013}.
Bilardi et al.~\cite{bilardi2000,bilardi2001characterization} develop
the notion of access complexity and relate it to space complexity.
Bilardi and Preparata \cite{Bilardi99} developed the notion of the
closed-dichotomy size of a DAG $G$ that is used to provide a lower bound
on the data access complexity in those cases where recomputation
is not allowed. Our notion of schedule wavefronts is similar
to the closed-dichotomy size in their work; but, unlike the work
of \cite{Bilardi99}, we use it do develop
an effective automated heuristic to compute lower bounds for CDAGs.
Extending the scope of the \hk model to more complex memory
hierarchies has been the subject of some research. Savage provided an
extension together with results for some classes of computations that
were considered by \hk, providing optimal lower bounds for I/O with
memory hierarchies \cite{savage.cc.95}. Valiant proposed a
hierarchical computational model \cite{valiant.jcss.11} that offers
the possibility to reason in an arbitrarily complex parameterized
memory hierarchy model.  

Unlike \hk's original model, several models have been proposed that do
not allow recomputation of values (also referred to as ``no
repebbling'')
\cite{BDHS11,BDHS11a,bilardi2001characterization,michele13,ranjan11.fft,savage.cc.95,savage.book,savage.options,cook.pebbling,toledo.jpdc,ranjan12.rpyr,ranjan12.vertex}.
Savage \cite{savage.cc.95} develops results for FFT using no
repebbling.  Bilardi and Peserico \cite{bilardi2001characterization}
explore the possibility of coding a given algorithm so that it is
efficiently portable across machines with different hierarchical
memory systems,  without the use of recomputation.  Ballard et
al.~\cite{BDHS11,BDHS11a} assume no recomputation is allowed in deriving lower
bounds for linear algebra computations.  Ranjan et
al.~\cite{ranjan11.fft} develop better bounds than \hk for FFT using a
specialized technique adapted for FFT-style computations on memory
hierarchies. Ranjan et al.~\cite{ranjan12.rpyr} derive lower bounds
for pebbling r-pyramids under the assumption that there is no
recomputation.
Recently, Ranjan et al.~\cite{ranjan12.vertex} develop a
technique for binomial graphs. 
\comment{
The use of \hk's model has required manual algorithm-specific reasoning to
find $S$-partitions, even for regular graphs. Savage's
\cite{savage.cc.95} $S$-span model also requires problem-specific
insights. We note that \hk's model and Savage's model are not strictly
comparable, as noted by Bilardi et al.~\cite{bilardi2000}.  The recent
works of Ranjan et
al.~\cite{ranjan11.fft,ranjan12.rpyr,ranjan12.vertex} develop a
technique inspired by FFT-style computations, r-pyramids and binomial
graphs. In addition to these works, other approaches
\cite{aggarwal.ca.88,DemmelGHL12,valiant.jcss.11,BDHS11a,bilardi2001characterization}
also require problem-specific insights to develop bounds.  
}
Very recent work from U.C. Berkeley
\cite{Demmel2013TR} has developed a very novel approach to developing
parametric I/O lower bounds 
applicable/effective for a class of nested
loop computations but is either inapplicable or produces weak lower
bounds for other computations (e.g., stencil computations, FFT, etc.).

The P-RBW game developed in this paper extends the parallel model for
shared-memory architectures 
by Savage and Zubair \cite{savage2008unified} to also include 
the distributed-memory parallelism present in all scalable parallel 
architectures. The works of Irony et al. \cite{toledo.jpdc} and 
Ballard et al. \cite{BDHS11} model communication across nodes of
a distributed-memory system. 
Bilardi and Preperata \cite{Bilardi99} develop lower bound results
for communication in a distributed-memory model specialized
for multi-dimensional mesh topologies. 
Our model in this paper differs
from the above efforts in defining a new integrated pebble game to model both
horizontal communication across nodes in a parallel machine, as well
as vertical data movement through a multi-level shared cache hierarchy
within a multi-core node.

Czechowski et al. \cite{czechowski2011balance,czechowski2013codesign} 
consider the relationship
between the ratio of an algorithm's data movement cost to arithmetic
work and the machine balance ratio of memory bandwidth to peak performance.
Our analysis of algorithms in this paper involves a very similar
theme as theirs and is inspired by their work, but we develop new
lower bounds analysis to perform the analysis. Further, we compare
and contrast data movement demands for horizontal across-node communication
versus vertical within-node data movement and observe that the latter is
often the more constraining factor.

\section{Conclusion}
\label{sec:conclusion}

Characterizing the parallel data movement complexity of a program 
is a cornerstone
problem, that is particularly important with current and emerging
power-constrained architectures where the data transfer cost
will be the dominant energy and performance bottleneck. In this paper
we presented an extension to the Hong and Kung red-blue pebble game
model to enable development of lower bounds on data movement
for parallel execution of CDAGs. The model distinguishes horizontal
data movement between nodes of a distributed-memory parallel system
from vertical data movement within the multi-level memory/cache hierarchy
within a multi-core node. The utility of the model and the 
developed lower bounding techniques was demonstrated by analysis of 
several numerical algorithms and the garnering of interesting insights
on the relative significance of horizontal versus vertical data movement
for different algorithms.

\bibliographystyle{abbrv}
\bibliography{iobib}

\end{document}